\documentclass[runningheads]{llncs}
\usepackage{cite}
\usepackage{graphicx}
\usepackage[utf8]{inputenc}
\usepackage{amsmath}
\usepackage{hyperref}
\usepackage{booktabs}
\usepackage{comment}
\begin{document}
\title{Improving self-supervised pretraining models for epileptic seizure detection from EEG data}


%
%
\author{Sudip Das \inst{1} \and
Pankaj Pandey \inst{1}\and
Krishna Prasad Miyapuram \inst{1}}
\institute{
Computer Science and Engineering, IIT Gandhinagar, India}
\authorrunning{Das, Pandey and Miyapuram}
%
%
\maketitle              
\begin{abstract}
There is abundant medical data on the internet, most of which are unlabeled. Traditional supervised learning algorithms are often limited by the amount of labeled data, especially in the medical domain, where labeling is costly in terms of human processing and specialized experts needed to label them. They are also prone to human error and biased as a select few expert annotators label them. These issues are mitigated by Self-supervision, where we generate pseudo-labels from unlabelled data by seeing the data itself. This paper presents various self-supervision strategies to enhance the performance of a time-series based Diffusion convolution recurrent neural network (DCRNN) model. The learned weights in the self-supervision pretraining phase can be transferred to the supervised training phase to boost the model's prediction capability. Our techniques are tested on an extension of a Diffusion Convolutional Recurrent Neural network (DCRNN) model, an RNN with graph diffusion convolutions, which models the spatiotemporal dependencies present in EEG signals. When the learned weights from the pretraining stage are transferred to a DCRNN model to determine whether an EEG time window has a characteristic seizure signal associated with it, our method yields an AUROC score $1.56\%$ than the current state-of-the-art models on the TUH EEG seizure corpus.

.

\end{abstract}
\section{Introduction}
A seizure is the occurrence of an abnormal, intense discharge of a batch of cortical neurons. Epilepsy is a condition of the central nervous system characterized by repeated seizures. The distinction between epilepsy and seizure is important because of the fact their treatments are different \cite{bromfield2006}. For epilepsy diagnosis, at least two unprovoked seizures should happen in quick succession. The risk of premature death in people with epilepsy is about three times that of the general population \cite{WHO2019}. The EEG or electroencephalogram is an electrophysiological-based technique for measuring the electrical activity emerging from the human brain normally recorded at the brain scalp.

It is widely accepted that EEG originates from cortical pyramidal neurons \cite{avitan2009eeg} that are positioned perpendicular to the brain's surface. The EEG records the neural activity, which is the superimposition of inhibitory and excitatory postsynaptic potentials of a sizeable group of neurons discharging synchronously. Cortical neurons are the electrically excitable cells present in the central nervous system. Through a special type of connection known as synapses, the information is communicated and processed in these neurons by electrochemical signaling.

Scalp Electroencephalogram (EEG) is one of the main diagnostic tests in neurology for detecting epileptic seizures. Most patients demonstrate some characteristic aberrations during an epileptic seizure\cite{hopp2016}. A drawback in using EEG signals for clinical diagnosis is that the cerebral potential may change quite drastically by muscle movements or even the environment. The resultant EEG signal contains artifacts that make it harder to provide a diagnosis to the affected patients. Most of the medical data available publicly are unlabeled \cite{lecun_2019}. Many approaches like semi-supervised learning and self-supervised learning have used the unlabeled data for training.

\section{Self-supervision in the medical domain}
There is an abundance of medical data on the internet, most of them being unlabeled. Traditional machine learning (ML) algorithms use labeled data to train supervised models. These unlabeled data are of no use to supervised models.

\begin{figure}[htp]
	\centering
	\includegraphics[width=10cm]{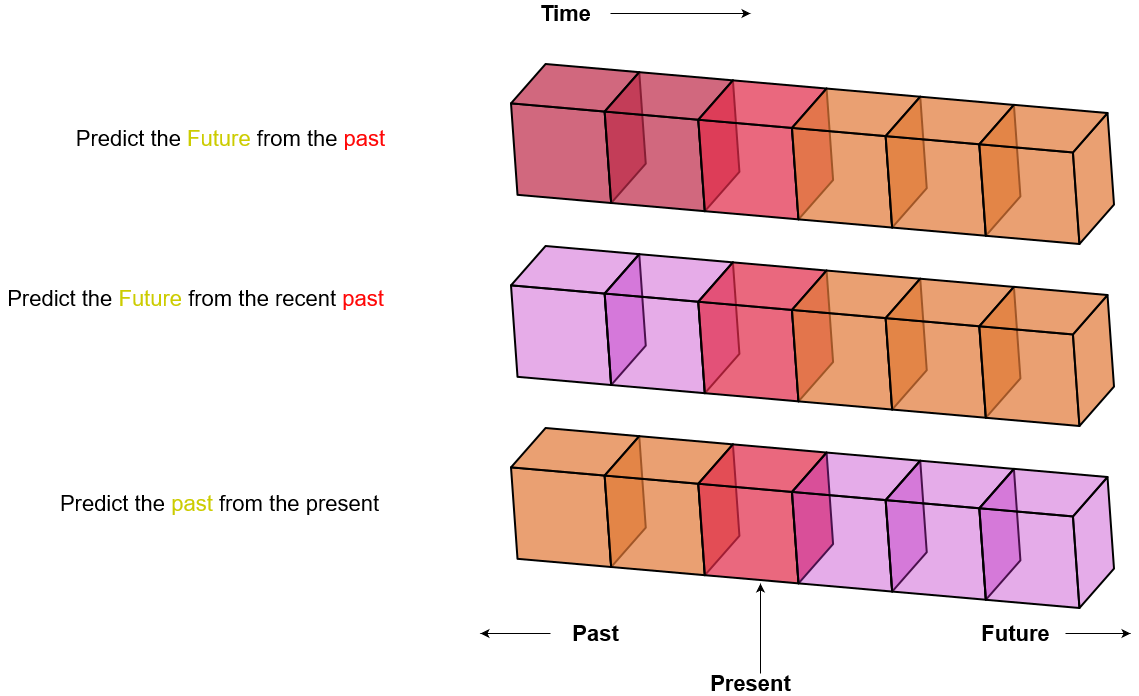}
	\caption{Basic pretext tasks for time series data}
	\label{abs}
\end{figure}

Self-supervised learning (SSL) methods, as the name suggests, use some form of supervision. Using self-supervision, the labels are extracted from the data by using a fixed pretext task. The task makes use of the innate structure of the input data \cite{doersch2015unsupervised}. Some of the basic pretext tasks are given in figure \ref{abs}. This strategy works well on images, as was proved in the ImageNet challenge \cite{russakovsky2015imagenet}. Self-supervision models such as SimCLR \cite{chen2020simple} and BYOL \cite{grill2020bootstrap} perform well in image classification tasks even if they use $1\%$ of the ground truth labels. They give performance comparable to the fully self-supervision tasks. It also works well on time series data such as natural language \cite{lan2019albert}, audio and speech understanding \cite{bai2020representation}.

In the Works of \cite{arora2019theoretical, wei2020theoretical, wei2020theoretical}, they tried to prove the usefulness of self-supervision and the qualities of a reliable self-supervision pretext task. In predictive learning, the representations of unlabeled data are learned by predicting masked time series data. Recent works \cite{saunshi2020mathematical} on time series data have shown how a contrastive learning technique can be reduced to a masking problem. However, most of these works are focused on computer vision or text-related applications, and very few of them use EEG data.

In this work, we use different self-supervised strategies where we use the unlabelled data to increase the performance of supervised models. These strategies help the model learn natural features from the dataset, which helps the model distinguish diseased Electroencephalogram (EEG) signals from the healthy ones.

\section{Related works}
There has been a shift from using convolutional neural networks (CNNs) to combining CNN and Graph neural networks. Many studies in this domain \cite{rasheed2020machine, raghu2020eeg,asif2020seizurenet,ahmedt2020identification}, assume euclidean structures in EEG signals. The application of 3D-CNN on the EEG spectrogram of the signals violates the natural geometry of EEG electrodes and the connectivity in the brain signals. EEG electrodes are placed on a person's scalp and hence are non-euclidean. The natural geometry of EEG electrodes is best represented on a graph \cite{bronstein2017geometric,chami2020machine}. In \cite{wang2020sequential}, the authors use GNN to extract feature vectors and then use convolution on the intermediate results. Whereas in \cite{wagh2020eeg} they extract features using Power Spectral Density (PSD) which are used as the node vectors, and perform spectral graph convolutions \cite{kipf2016semi}. Graph representation learning has been used heavily in modeling brain networks \cite{bullmore2009complex}. Recent models such as \cite{tang2021self} outperforms previous CNN based methods \cite{saab2020weak} and CNN based Long short-term memory-based methods \cite{ahmedt2020identification}. In \cite{tang2021self}, they use self-supervision to increase the performance of the supervision models. Specifically, they pre-train the model on the $12/60s$ of EEG clip for predicting the next $T^{'}$ second  EEG clip and transfer the learning to the model's encoder. Some studies \cite{xu2020anomaly, banville2021uncovering , kostas2021bendr, martini2021deep} used a different pretraining strategy but did not use any graph-based methods to model the EEG signals.

\section{Experimental setup}

\subsection{Dataset}
For our testing, we used the publicly available Temple university hospital EEG Seizure corpus (TUSZ) v1.5.2. \cite{shah2018temple}, which is the largest publicly available seizure corpus to date with $5,612$ EDF files, $3,050$ annotated seizures from clinical recordings, and eight seizure types. Five subjects were present both in the TUSZ train and test subjects, thus not considered in our self-supervision strategy. The validation split was created from the train split by randomly selecting $10\%$ of the subjects from the train split. This was done to ensure the test and validation performance reflects the performance of unseen subjects as seen in real-world scenarios. 

Since epileptic seizures have a specific frequency range in the EEG signal. \cite{tzallas2009epileptic}, and hence it is intuitive to convert the raw EDF files into the frequency domain. The log amplitudes are obtained after applying the Fast Fourier transformation to all the non-overlapping $12s$ windows. This is done for all the windows for faster pre-processing. 

\begin{table}[]
\centering
\begin{tabular}{||c|cccc||}
\hline
 &
  \multicolumn{1}{l}{\textbf{\begin{tabular}[c]{@{}l@{}}EEG files \\ (\% Seizure)\end{tabular}}} &
  \multicolumn{1}{l}{\textbf{\begin{tabular}[c]{@{}l@{}}Patients \\ (\% Seizure)\end{tabular}}} &
  \multicolumn{1}{l}{\textbf{\begin{tabular}[c]{@{}l@{}}Total \\ Duration \\ (Sec)\end{tabular}}} &
  \multicolumn{1}{l||}{\textbf{\begin{tabular}[c]{@{}l@{}}Seizure \\ Duration\\ (Sec)\end{tabular}}} \\ \hline
\textbf{Train files} &
  4599 (18.9\%) &
  4599 (34.1\%) &
  2710483 &
  169793 (6.6\%) \\ \hline
\textbf{Test files} &
  900 (25.6\%) &
  45 (77.8\%) &
  541895 &
  53105 (9.8\%) \\ \hline
\end{tabular}
\caption{Detailed summary of TUSZ v1.5.2}
\label{T-1}
\end{table}

\section{Graph neural network for EEG signals}
\subsection{Representing EEG's as graphs}
Multivariate time-series of EEG signals can be modeled as a graph $G$, where $G = {V, E, W}$. Here $V$ denotes the electrodes/channels, $E$ denotes the set of edges, and $W$ is the adjacency matrix. Two types of graphs were formed according to  \cite{tang2021self} to validate our self-supervision strategies.

\textbf{Distance Graph:} To capture the natural geometry of the brain, edge weights are computed between two electrodes $v_i$ and $v_j$ is the distance between them, i.e. 
\begin{equation}
    W_{ij} = exp(-\frac{dist(v_i, v_j)^2}{{\sigma}^2}) \hspace{15pt} if \hspace{5pt}dist(v_i, v_j) \leq \kappa
\end{equation}

Where $\kappa$ is the threshold produced by a Gaussian kernel \cite{shuman2013emerging}, which generates a sparse adjacency matrix, and $\sigma$ is the standard deviation of the distances. This results in a weighted undirected graph for the EEG windows. The $\kappa$ is set at $0.9$ as it resembles the EEG montage widely used in the medical domain \cite{acharya2016american}.

\textbf{Correlation graph:} The edge weight $W_{ij}$ is computed as the absolute value of the cross-correlation between the electrode signals $v_{i}$ and $v_{j}$, normalized across the graph. To keep the more influential edges and introduce sparsity, the top $3$ neighbors for each node are kept, excluding self-edges.

\subsection{DCRNN for EEG signals}
A Diffusion Convolutional Recurrent neural network (DCRNN)\cite{li2017diffusion} models the \textit{spatiotemporal dependencies} in EEG signals. DCRNN was initially proposed for traffic forecasting on road networks, where it is modeled as a diffusion process. The objective is to learn a function that predicts the future traffic speeds of a sensor network given the historic traffic speeds.

The spatial dependency of the EEG signals is similar to the traffic forecasting problem and is modeled as a diffusion process on a directed graph. This is because an electrode has an influence over other electrodes given by the edge weights. DCRNN captures spatial and temporal dependencies among time series using
diffusion convolution, sequence to sequence learning framework, and scheduled sampling.

\textbf{Diffusion Convolution} 
The diffusion process is characterized by a bidirectional random walk on a directed graph signal $G$. Whereas the diffusion convolution operation over a graph signal $X \in R ^{N\times P}$ and a filter $f_\theta$ is defined as:
\begin{equation}
    X_{:,p\  \star G \ f_\theta} = \sum_{k=0}^{K-1} (\theta_{k, 1}{(D_O^{-1} W})^k + \theta_{k, 2}{(D_I^{-1} W^{T}})^k)X_{:,p} \ \ \ for \ p \in \{1, \cdots, P\}
\end{equation}
where $X \in R^{N \times P}$ is the pre-processed EEG signal at a specific time instance $t \in \{1,\cdots\,T\}$ with N nodes and P features. Here $\theta \in R^{K \times 2}$ are the parameters of the filter and $D_O^{-1}W$, $D_I^{-1} W^T$ represent the transition matrices of the diffusion matrices of the inwards and outwards diffusion processes respectively\cite{li2017diffusion}.

For modeling the temporal dependency in EEG signals, Gated Recurrent units \cite{cho-etal-2014-properties}, a variant of Recurrent neural network (RNN) having a gating mechanism are used. To incorporate diffusion convolution, the matrix multiplications are replaced with diffusion convolutions \cite{li2017diffusion}. For seizure detection, the model consists of several DCGRUs, stacked together and accompanied by a fully connected layer.

\section{Self-Supervision Methods}
We propose five pretraining strategies for time series data for model pretraining. These methods are tested on a T = $12s$ window, having $19$ channels and $200$ time points. We denote the matrix corresponding to the EEG signal as $S \in R ^ {19 \times 200}$. The objective function is to predict the original signal given the noisy/masked signal. We use the mean absolute error between the predicted and the original EEG window as our loss function.

\subsection{Jitter}
\label{1}
\begin{figure}[htp]
    \centering
    \includegraphics[width=12cm]{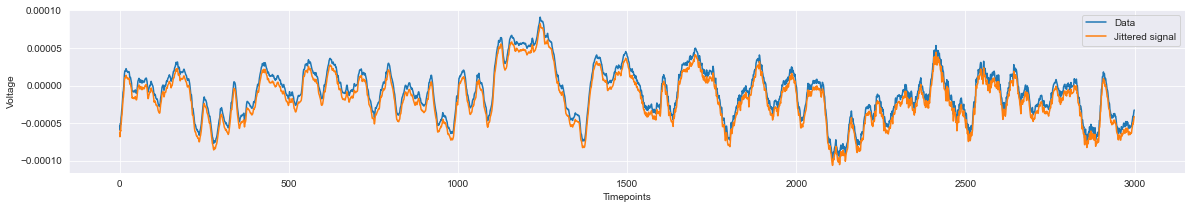}
    \caption{EEG signal before and after introducing jitter}
    \label{fig:I2}
\end{figure}

The modified signal, $S^{'}$ was created by superimposing the main signal, $S$ with a noise signal. The noise has the same mean, $\mu(S)$ and a fraction $L$ of the variance of the original signal. We set $L = 5\%$ as experimented by \cite{jayarathne2020person} for data augmentation. The random noise signal $M$ from a normal distribution, is created as $M=N(\mu(S),0.05 \cdot Var(S))$ . Then, 
\begin{equation}
    S^{'}=S+M
    \label{eq1}
\end{equation}
Since the magnitude of the noise is much lower compared to the EEG signal, the resultant signal had a small shift along the y-axis, as shown in Figure \ref{fig:I2}. 

\graphicspath{ {Images/} }

\subsection{Random sample}
\label{2}
\begin{figure}[htp]
    \centering
    \includegraphics[width=12cm]{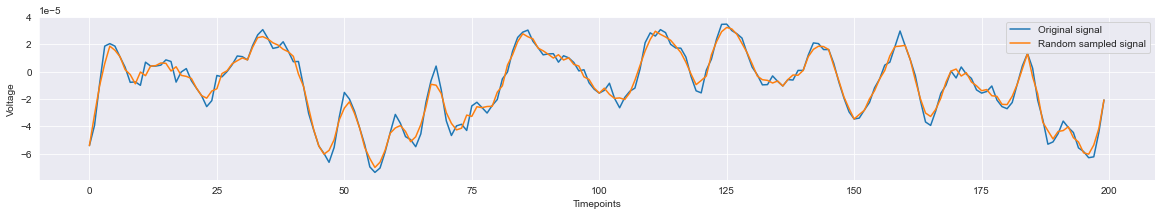}
    \caption{EEG signal before and after introducing random sampling}
    \label{fig:I3}
\end{figure}

The noisy signal is created by randomly choosing some fraction of the time points, i.e., $t_{1},t_{2}, \cdots t_{k}$ and replacing them with their neighbors' average\cite{jayarathne2020person}. For example, if neighbors of $t_i$ are $t_{i1},t_{i2}$ then we replace $t_i$ with $\frac{t_{i-1}+t_{i-1}}{2}$. The first and last points were not considered as they do not have both neighbors. We chose $20\%$ of the points in each channel and replaced them with zeros, as shown in Figure \ref{fig:I3}. This has a "smoothening" effect on the initial graph as some data points which were local maxima/minima are now in the straight line connecting their neighbors.

\subsection{Remove a specific channel} \label{3}
Trying to predict a channel by using the data from other channels clears the way for many different opportunities. It proves that a particular channel is a function of other channels and does not provide any new data.

\begin{figure}[htp]
    \centering
    \includegraphics[width=12cm]{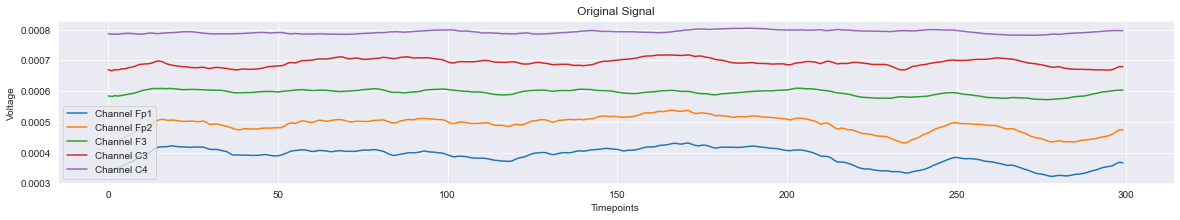}
    \caption{Original  signal with five channels}
    \label{fig:I4}
\end{figure}

We try to use this problem in our self-supervision. This can be done either by picking a random channel from each lobe or for all the channels. We select channel F3 from the Frontal lobe for this task. Mathematically, we set $S_{K, i}=0$ for all $i$, where $K$ is the channel number. For illustration purposes, Figure. \ref{fig:I4} only uses the first five channels in which the channel F3 is replaced with zeros. 

\begin{figure}[htp]
    \centering
    \includegraphics[width=12cm]{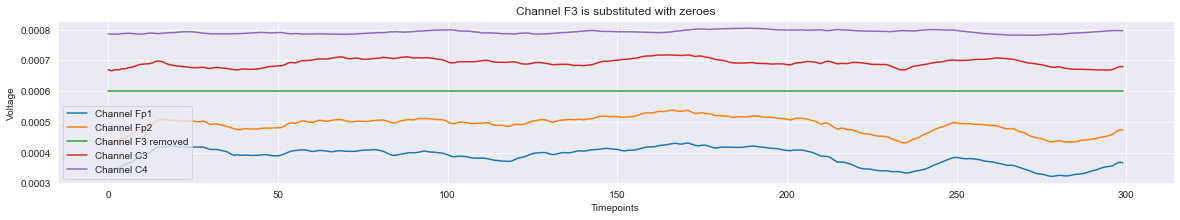}
    \caption{Resultant signals after noise is added}
    \label{fig:I5}
\end{figure}

\subsection{Predicting windows} \label{4}
We select a small window from each channel instead of selecting all the points of a single channel. The noisy signal is created by replacing the selected time points with dummy points for all the channels.

\begin{figure}[htp]
    \centering
    \includegraphics[width=12cm]{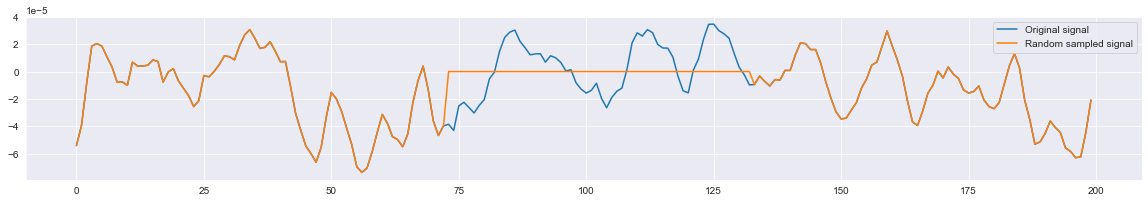}
    \caption{Resultant signal after noise is added for a single channel}
    \label{fig:I6}
\end{figure}

In other words, we choose a window of length $l$ randomly in each of the channels $i$ and make entries from $j+1$ to $j+l$ of that channel zero. So, for a particular channel $i$, $S_{i,j+1}, \cdots ,S_{i,j+l}= 0$. The window length $l$ consists of $20\%$ the total timepoints. These points are replaced with zeros. This is shown in Figure \ref{fig:I6}.

\subsection{Jittering in random window}
This method is a combination of methods \ref{1} and \ref{4}. A noise having the same characteristics as in \ref{1} is introduced in a small window having the same length as in \ref{4}. This method combines the advantages of both, as no essential data is being deleted but modified only in a selected window.
In other words, we choose a window of length $l$ randomly in each of the channels $i$. We generate a random noise $M$ with the same strategy as in \ref{1}, and introduce this noise in a window that is selected by \ref{4}. We use equation \eqref{eq1} for generating noise. Mathematically,
\begin{equation}
    S^{'}_{i,j+k}=S_{i,j+k}+ 0.05 \cdot Var(M)
    \ for\ k=1\ to\ l.
\end{equation}

\begin{figure}[htp]
    \centering
    \includegraphics[width=12cm]{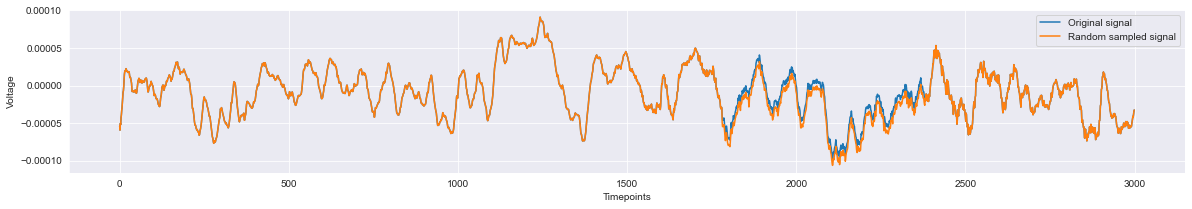}
    \caption{A window is selected and noise is introduced}
    \label{fig:I7}
\end{figure}
\begin{table}[t]
	\caption{Comparison of our proposed techniques with recent work}
	\centering
	\begin{tabular}{|l|l|} 
		\toprule
		\textbf{Seizure detection model}                        & \textbf{Seizure detection AUROC}  \\ 
		\hline
		\textbf{Traditional algorithms \cite{tang2021self}}                               &                          \\ 
		\hline
		Dense-CNN                    & 0.812 $\pm$0.014         \\
		LSTM                    & 0.786$\pm$ 0.014        \\
		CNN-LSTM                     & 0.749 $\pm$0.006         \\

		\hline
		\textbf{State of the art \cite{tang2021self}}                               &                          \\ 
		\hline
		Corr-DCRNN w/o Pre-training                    & 0.812 $\pm$0.012         \\
		Dist-DCRNN w/o Pre-training                    & 0.824 $\pm$ 0.020        \\
		Corr-DCRNN w/ Pre-training                     & 0.861 $\pm$0.005         \\
		Dist-DCRNN w/ Pre-training                    & 0.866 $\pm$0.016         \\
		
		\hline
		\multicolumn{1}{l}{\textbf{Our SSL strategies}}                       &                          \\ 
		\hline
		
		Corr-Jittering w/Pre-training & \textbf{0.8709 $\pm$ 0.0047}        \\
		Corr-Jittering in a random window w/pretraining                      & 0.8658 $\pm$  0.0271     \\
		Corr-Remove channel (F3) w/Pre-training        & 0.8655 $\pm$ 0.0035              \\
		Corr-Predicting windows w/Pre-training         & 0.8691 $\pm$ 0.0040        \\

		Corr-Random sample w/Pre-training              & 0.8692 $\pm$ 0.0042                    \\
		& \\
		\hline 
		Dist-Jittering w/Pre-training & 0.8749 $\pm$ 0.000053        \\
		Dist-Jittering in a random window                       & 0.8759 $\pm$  0.000045     \\
		Dist-Remove channel (F3) w/Pre-training        & 0.8774 $\pm$ 0.000035              \\
		Dist-Predicting windows w/Pre-training         & 0.8802 $\pm$ 0.000012          \\

		Dist-Random sample w/Pre-training              & \textbf{0.8816 $\pm$ 0.000006}                    \\
		
		\bottomrule
	\end{tabular}
	\label{table:Result}
\end{table}

In Figure \ref{fig:I7}, a certain part of the signal has some modifications, while the rest is left unchanged.

\section{Experimental results and analysis}

\subsection{Self-Supervision proposed techniques improves performance}
Self-supervision is typically used when we want to use unlabeled data for training our model. In contrast, the TUSZ dataset used in our study is fully labeled. In works of \cite{arora2019theoretical, wei2020theoretical, wei2020theoretical} they tried to prove why self-supervision helps and what are the qualities of a reliable self-supervision pretext task. The study which we extended \cite{tang2021self} also proves the same. We were able to improve the gap by using different pretraining strategies, especially on the distance-based graphs.

\subsection{Model training}

The AUROC score is the industry standard for evaluating models in the medical domain. It is prevalent for binary classification in medical domain \cite{ahmedt2020identification, o2020neonatal, asif2020seizurenet}. Therefore, we use the AUROC score as the standard evaluation metric for evaluating our self-supervision strategies. The base model of predicting the next $T = 12 s$, given the current $T = 12 s$ in \cite{tang2021self} has been replaced with these strategies. The batch size has been set to $1500$, while the other hyperparameters have been set at the default values provided by \cite{tang2021self}. Our pretraining and training were performed on a single NVIDIA RTX 5000 GPU. The self-supervision model parameters were randomly initialized for the pretraining phase. The DCGRU decoder model parameters were randomly initialized, and the encoder weights were set to the same weights as that of the self-supervision encoder.

\subsection{Outperform the state-of-the-art}
We compare our five self-supervision models with the previous state-of-the-art in Table \ref{table:Result}. For each strategy and type of graph, we ran the pretraining twice. We took five training runs for each self-supervision strategy to get a reliable estimate. The mean and standard deviation are computed across all the models.

The self-supervision model weights are transferred to the encoder of the seizure classification model for each run and then trained on the labeled windows. With these changes applied, we were able to surpass the state-of-the-art results by $1.56\%$ after applying this simple modification.

\subsection{Low deviation during Dist-Graph Compared to Correlation Graph}
The correlation-based graphs either produced very marginal improvement or had a lower score compared to the state-of-the-art. We set the number of epochs for pretraining and training as $350$ and $100$, respectively. The loss function for the distance-based graph converged in both the training and testing phases over all the epochs. In contrast, the loss function for correlation-based graphs converged faster at a higher value than their distance-based counterparts. The correlation-based models were stopped early since we did not see any improvements. This happened during both the training and testing phases. This resulted in a higher standard deviation and a lower AUROC score than their respective distance-based counterparts.

We noted that in papers \cite{song2018eeg, tang2021self, wang2018eeg, varatharajah2017eeg}, they used spatial connectivity as well as distance functional connectivity. For spatial connectivity, they used a distance-based graph, either the Euclidean distance or the actual measured between two points of a sphere. For functional connectivity, they have used a binary relationship between the electrodes. The purpose of using functional connectivity is to exploit the dynamic nature of time-series data. To validate our results, we randomly re-ran two of these five strategies for both types of graphs, and our results were within the margin of error.

\section{Conclusion and future work}
A notable point in our results is the difference in the magnitude of the standard deviation. The order of magnitude of the correlation-based models is similar to that of the previous state-of-the-art models. However, it is considerably more precise in distance-based models. The distance graphs consistently outperformed the correlation-based graphs and benefitted the most after using these five SSL strategies. The distance-based random sampling model performed the best among all the models, whereas the correlation-based random sampling model came in a close second among the correlation-based ones. We outperformed the previous state-of-the-art by introducing the self-supervision strategy, and the training phase was kept as it is. Our future work will focus on testing our proposed novel self-supervision strategies to other time-series data.

\section*{Acknowledgement}
We thank Science and Engineering Research Board (SERB) and PlayPower Labs for supporting the Prime Minister’s Research Fellowship (PMRF) awarded to Pankaj Pandey. We thank Federation of Indian Chambers of Commerce \& Industry (FICCI) for facilitating this PMRF.

\bibliographystyle{splncs04}
\bibliography{refs}
\end{document}